\pdfoutput=1
\documentclass[aps,prstab,twocolumn,groupedaddress,nofootinbib,floatfix]{revtex4-1}

\usepackage{graphicx}
\usepackage{booktabs}
\usepackage[caption=false]{subfig}
\usepackage{amsmath}
\usepackage{multirow}

\newcommand{\mbf}[1]{\mathbf{#1}}
\renewcommand{\(}{\left(}
\renewcommand{\)}{\right)}

\mathchardef\mhyphen="2D

\begin{document}
\title{Measurement and Compensation of Horizontal Crabbing at the 
Cornell Electron Storage Ring Test Accelerator}

\author{M. P. Ehrlichman}
\email[]{mpe5@cornell.edu}
\author{A. Chatterjee}
\author{W. Hartung}
\author{B. Heltsley}
\author{D. P. Peterson}
\author{N. Rider}
\author{D. Rubin}
\author{D. Sagan}
\author{J. P. Shanks}
\author{S. T. Wang}
\affiliation{Cornell Laboratory for Accelerator-based Sciences and 
Education, Cornell University, Ithaca, NY}

\begin{abstract}
In storage rings, horizontal dispersion in the rf cavities introduces 
horizontal-longitudinal ($xz$) coupling, contributing to beam tilt 
in the $xz$ plane.  This coupling can be characterized by a ``crabbing'' 
dispersion term $\zeta_a$ that appears in the normal mode decomposition
of the $1$-turn transfer 
matrix.  $\zeta_a$ is proportional to the rf cavity voltage and the 
horizontal dispersion in the cavity. 
We report experiments at the Cornell Electron Storage Ring Test Accelerator
(CesrTA) where $xz$ coupling was explored using three lattices with distinct 
crabbing properties. We characterize the $xz$ coupling for each case by 
measuring the horizontal projection of the beam with a beam size monitor.
The three lattice configurations correspond to a) $16$ mrad $xz$ tilt at 
the beam size monitor source point, b) compensation of the $\zeta_a$ 
introduced by one of two pairs of RF cavities with the second, and c) zero 
dispersion in RF cavities, eliminating $\zeta_a$ entirely.  Additionally, 
intrabeam scattering (IBS) is evident in our measurements of beam size vs. 
rf voltage.
\end{abstract}

\maketitle

\section{INTRODUCTION}
\label{sec:intro}
Just as coupling of horizontal and vertical motion can result in a bunch 
profile that is tilted in the transverse plane, coupling of horizontal and 
longitudinal motion will in general produce a tilt in the horizontal-longitudinal 
($xz$) plane.  The requisite $xz$ coupling can be generated by dispersion in 
ordinary rf accelerating cavities.

Crabbing in electron storage rings has been explored at 
KEKB \cite{kekb-crab:07}.  There, specially constructed crab cavities were 
used to generate an $xz$ tilt of $22$ mrad at the interaction region in order 
to compensate for the crossing angle.  Tilt angles at KEKB were measured 
directly using a streak camera.  

\begin{figure}
   \centering
   \includegraphics*[width=1.00\columnwidth]
                  {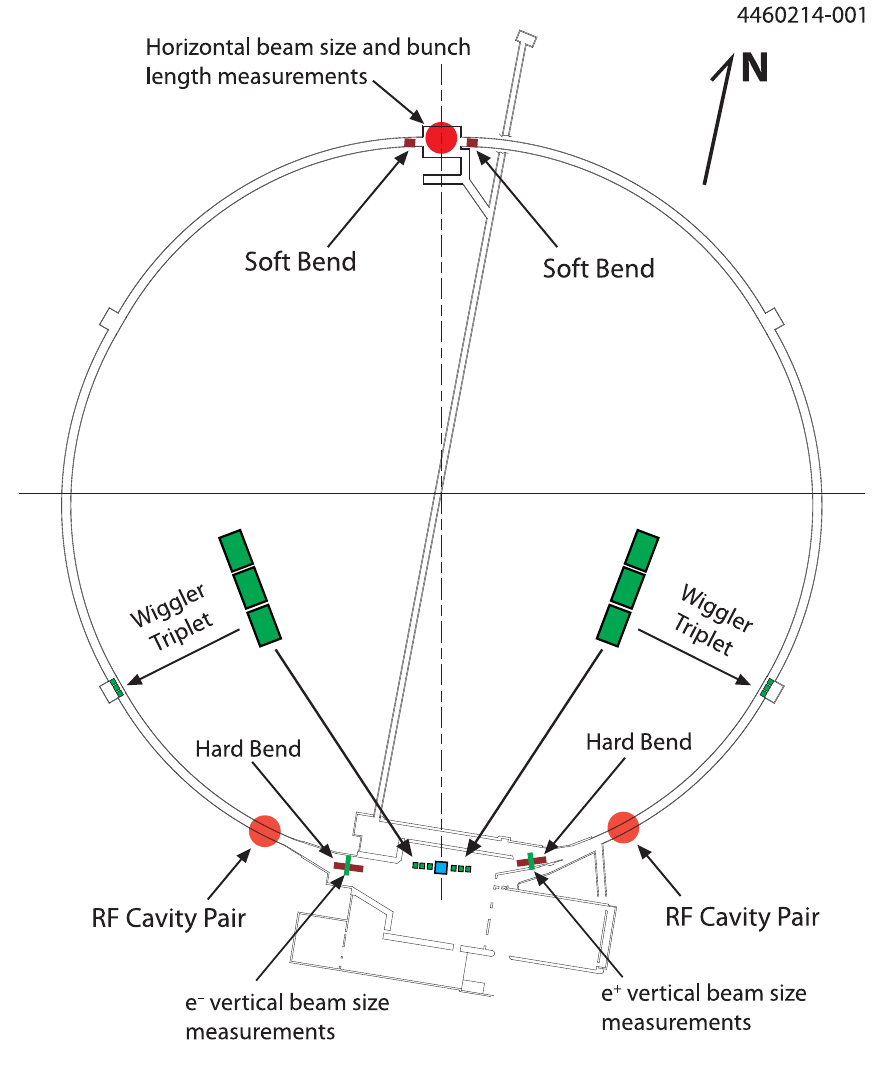}
   \caption{The configuration of the
   Cornell Electron Storage Ring Test Accelerator \cite{cesrta:2009}.
   \label{fig:cesrta}}
\end{figure}

Studies of $xz$ tilt have been done at the Cornell Electron Storage
Ring Test Accelerator (CesrTA).  At CesrTA, we measure tilt indirectly by 
observing the horizontal projection of the beam as rf voltage is varied.
Some relevant machine parameters are shown in Table \ref{tab:cesrta}.
The CesrTA layout is shown in Fig.~\ref{fig:cesrta}.
\begin{table}[b]
\centering
\caption{CesrTA machine parameters for crabbing studies.\label{tab:cesrta}
$\nu_{x,y,s}=2\pi Q_{x,y,s}$ is used in this paper.}
\begin{tabular*}{\columnwidth}{@{\extracolsep{\fill}}lcr}
\hline
\hline
Beam Energy (GeV)             &$E_0$              & $2.085$\\
Circumference (m)             &$L$                &$768$\\
RF Frequency (MHz)            &$\omega_{rf}$      &$2\pi\times 500$\\
Transverse Damping Time (ms)  &$\tau$             & $56.6$\\
Momentum Compaction           &$\alpha_p$         & $0.0068$\\
Nominal RF Voltage (MV)       &$V_{rf}$           & $6.3$\\
Synchrotron tune              &$Q_s$              & $0.065$\\
Horizontal tune               &$Q_x$              & $14.624$\\
Vertical Tune                 &$Q_y$              & $9.590$\\
Horizontal Emittance (nm\textperiodcentered rad)  &$\epsilon_a$     &$\sim3$\\
Vertical Emittance (pm\textperiodcentered rad)    &$\epsilon_b$     &$\sim5$ to $\sim15$\\
\hline
\end{tabular*}
\end{table}
The horizontal beam size monitor source point, the two rf cavity pairs, and 
the four damping wiggler triplets are highlighted.  The rf straights are in close 
proximity to the interaction region/damping wiggler straight in the South arc.
Because of the intervening hard bend 
magnets, there is no practical lattice solution with zero dispersion in both 
the rf and wiggler straights.  
In order to minimize the horizontal emittance, we generally opt 
for zero dispersion in the wiggler straight.  The result is horizontal
dispersion of about $1$ meter in the rf cavities.  This horizontal dispersion 
results in a tilt of the beam in the $xz$ plane by an amount that depends on 
the total rf accelerating voltage.

The horizontal beam size monitor at CesrTA measures the projection of the 
beam into the horizontal lab frame coordinate.  A tilt of the beam in the $xz$ 
plane manifests itself as an increase in the measured horizontal beam size.
Bunch lengths in CesrTA are typically $10$ mm and bunch widths are 
typically $150$ $\mu$m.  Even small amounts of tilt can result in a 
significantly larger measured horizontal size.

The largest inferred tilt at the beam size monitor source point is $16$ mrad.  
The tilt is modulated by the horizontal phase advance, and the largest
tilt in our lattice model is $47$ mrad.

In this paper, we describe recent experiments at CesrTA to measure and 
correct the $xz$ tilt, and the theoretical basis for our
correction techniques.
The calculations presented here were conducted using the {\tt BMAD} 
accelerator simulation suite \cite{bmad:2006}.

Adjusting the rf voltage also changes the bunch length and hence
the particle density, which in turn 
changes the amount of IBS blow up.  In addition to measuring the effect of 
beam tilt, we observe IBS effects, as we measure vertical as well
as horizontal size.

In Sec.~\ref{sec:toy} we show with a simple model how dispersion in the rf
cavities leads to a beam that is tilted in the $xz$ plane.

In Sec.~\ref{sec:theory}, we develop a parameterization of the one-turn
matrix and a numerical method for obtaining it.
From this parameterization, we extract the tilt and 
projected size of beams in arbitrary coupling conditions.  
This section's intent is to give the reader a more rigorous and complete 
picture of the parameterization and to explain how it describes the 
beam envelope.

In Sec.~\ref{sec:lats}, we present two methods for eliminating the tilt:  
(i) canceling the crabbing dispersion $\zeta_a$ by adjusting the horizontal 
phase advance between 
the two pairs of rf cavities;  (ii) constraining the optics so that the 
dispersion is zero in the RF straights.  This necessarily results in non-zero 
dispersion in the wiggler straight, hence for operation in these optics, $6$ 
of the $12$ damping wigglers must be turned off.

In Sec.~\ref{sec:experiment}, we test our formalism for calculating beam 
sizes, as well as our tilt-mitigating lattices, by measuring the beam size as 
a function of rf voltage.

Manipulation of $xz$ coupling can be useful for both collider damping rings
and light sources.  
In a collider, $xz$ coupling can be used to achieve full head-on collisions.
The technique is being considered as part of a future LHC 
luminosity upgrade \cite{lhc-cc-ipac}.  In light sources,
$xz$ coupled beams have been proposed as a method for obtaining 
subpicosecond x-ray pulses from a storage ring \cite{aps-crabbing}.

Typically, $xz$ coupling is obtained using crab cavities.  Crab
cavities use the magnetic field of a TM$_{\mathrm{110}}$ horizontal
dipole mode to apply a phase-dependent transverse kick to the beam.
The formalism presented here is a potentially simpler
way to manipulate $xz$ coupling in an accelerator.  It does
not require new technology and is based on elementary beam optics.

\section{Brief Description of Crabbing Due to Dispersion in an rf Cavity}
\label{sec:toy}
Rf cavities are present in a storage ring to restore energy to the beam that 
is lost due to synchrotron radiation.  
In machines operating above transition, the rf phase is set so that
the lower-energy particles which arrive at the rf cavity sooner
receive a bigger energy kick, while the higher-energy particles which
arrive later receive a smaller kick.
The kick received is
\begin{equation}
\frac{\Delta E}{E_0}=\frac{e V_{\mathrm{rf}}}{E_0}\sin\(\Psi_0-\Delta\Psi\),
\label{eqn:toy-DE}
\end{equation}
where $E_0$ is the beam energy, $e$ is the electric charge, 
$V_{\mathrm{rf}}$ is the peak cavity
voltage, $\Psi_0$ is the nominal synchrotron phase, and
\begin{equation}
\Delta\Psi=\frac{z\omega_{\mathrm{rf}}}{\beta_r c},
\label{eqn:toy-DP}
\end{equation}
where $z$ is the longitudinal coordinate relative
to the reference particle, $\omega_{\mathrm{rf}}$ is the rf frequency,
$c$ is the speed of light, and $\beta_r$ is the relativistic beta. 
Particles at the head of the bunch have 
a positive $z$.  Above transition,
$\frac{\pi}{2}<\Psi_0<\frac{3\pi}{2}$, and typically $\Psi_0\sim\pi$. 
Expanding Eq.~(\ref{eqn:toy-DE})
about $\Psi_0=\pi$ yields,
\begin{equation}
\frac{\Delta E}{E_0}\approx \tilde{V}z
\label{eqn:toy-DEapprox}
\end{equation}
where 
\begin{equation}
\tilde{V}\equiv\frac{eV_{rf}\omega_{rf}}{\beta_r cE_0}.
\label{eqn:Vtilde}
\end{equation}

The position $x$ of a particle has contributions from betatron motion
and dispersion $\eta_x$,
\begin{equation}
x=x_\beta+\delta\eta_x,
\end{equation}
where $\delta$ is the relative energy deviation of the particle.
In a ``zero-length'' cavity, a particle receives an energy kick 
$\Delta E$, but its instantaneous position does not change.  So,
\begin{equation}
\Delta x=0=\Delta x_\beta+\frac{\Delta E}{E_0}\eta_x \rightarrow \Delta x_\beta=
-\frac{\Delta E}{E_0}\eta_x.
\label{eqn:betatron}
\end{equation}

Suppose that $\eta_x\neq 0$ in the cavity.  The formula for the 
change in the closed orbit at location $s$ due to a displacement 
$\Delta x_0$ at location $s_0$ is
\begin{equation}
x\(s\)=-\Delta x_0\sqrt{\beta_x\(s\)\beta_{x0}} \frac{
\gamma_{x0}\sin\(\Delta\phi\(s\)-\pi Q_x+\phi_{\alpha0}\)
}{2\sin\pi Q_x},
\label{eqn:x-kick}
\end{equation}
where $s$ is the location of the reference particle in the accelerator,
$\Delta\phi\(s\)$ is the phase advance from $s_0$ to $s$, 
$\beta_x\(s\)$ is the horizontal Twiss parameter at $s$, 
$\phi_{\alpha0}=\textrm{arcsin}\frac{-\alpha_{x0}}{\sqrt{1+\alpha_{x0}^2}}$, 
and $\beta_{x0}$, $\alpha_{x0}$, and $\gamma_{x0}$ are the horizontal 
Twiss parameters at $s=s_0$.

Setting $\Delta x_\beta=\Delta x_0$ and $\eta_x=\eta_{x0}$ in 
Eq.~(\ref{eqn:betatron}), and combining Eqs.~(\ref{eqn:toy-DEapprox}),
(\ref{eqn:betatron}), and (\ref{eqn:x-kick}), 
and dividing by $z$ yields the $xz$-tilt angle,
\begin{multline}
\theta_{xz}\(s\)=\tan\frac{x\(s\)}{z}\approx\\
\tilde{V}\eta_{x0}\sqrt{\beta_x\(s\)\beta_{x0}} \frac{
\gamma_{x0}\sin\(\Delta\phi\(s\)-\pi Q_x+\phi_{\alpha0}\)
}{2\sin\pi Q_x}.
\label{eqn:theta-by-pert}
\end{multline}

Equation~(\ref{eqn:theta-by-pert}) gives the $xz$ tilt at some location
$s$ due to dispersion in an rf cavity at $s_0$.  A similar treatment
would reveal the tilt due to having finite $\eta'_x$ in an rf cavity.

Inspecting Eq.~(\ref{eqn:theta-by-pert}) we see that if one were to 
follow the beam around the ring, the tilt angle would be 
observed to oscillate as the betatron phase advances, and the amount of
tilt is proportional to $\sqrt{\beta_x}$ at the observation point.

Later, in Sec.~\ref{sec:simplified}, 
equations are
derived for the beam tilt in terms of the parameterization of the one-turn
matrix.  Equation (\ref{eqn:theta-by-pert}) agrees with 
these later results in the appropriate limits.

\section{THEORY}\label{sec:theory}
Consider the one-turn map $\mbf{T}_4$ for four-dimensional phase 
space $\(x,p_x,y,p_y\)$.  The Jacobian of the map is a symplectic
$4\times 4$ matrix.  The Edwards-Teng parameterization expresses $\mbf{T}_4$ 
in terms of $10$ parameters \cite{edwards-teng}:  two normal-mode 
phase advances, four normal-mode Twiss parameters, and four coupling parameters.
The coupling parameters describe how the normal-mode coordinates 
$\(a,p_a,b,p_b\)$ transform into lab-frame coordinates $\(x,p_x,y,p_y\)$.  
Three of the four coupling parameters can be measured directly in CESR, 
which allows 
for optics correction \cite{SaganRubin}.  Additionally, in the limit of linear 
optics and Gaussian beams, normal-mode emittances are well-defined invariants.

The Edwards-Teng parameterization is incomplete in that it ignores longitudinal 
motion.  In \cite{ohmi:1994}, Ohmi, Hirata, and Oide extend the Edwards-Teng 
parameterization to the full $6\times6$ transfer matrix $\mbf{T}$.  $\mbf{T}$ 
is the one-turn map for six-dimensional phase space $\(x,p_x,y,p_y,z,p_z\)$.  
It is described by $21$ parameters: three normal-mode phase advances, six 
normal-mode Twiss parameters, and twelve coupling parameters.

The $6\times6$ parameterization of the one-turn map is useful when there is 
significant coupling between the longitudinal motion and the transverse
motion.  The orientation of the beam envelope can be written in terms
of the coupling parameters.  With this description, the coupling properties 
of the ring can be adjusted or corrected by varying these parameters.  
Additionally, the normal-mode Twiss parameters $\beta_a$, $\alpha_a$, 
$\beta_b$, $\alpha_b$, $\beta_c$, and $\alpha_c$, and normal-mode emittances 
$\epsilon_a$, $\epsilon_b$, and $\epsilon_c$ are well-defined quantities.

In this section we present the $6\times6$ parameterization in a format that 
is convenient for investigating tilt in the $xz$ plane, and we also describe
a numerical method for obtaining it.  Our description 
differs from that in \cite{ohmi:1994} in that we only use real-valued 
quantities.  We extend the formalism by defining ``normalized'' 
coupling parameters which simplify the expressions for the beam tilt and 
beam size.

\subsection{The $6\times6$ parameterization}
The Edwards-Teng parameterization is extended to
the $6\times6$ case via the ``dispersion matrix'' $\mbf{H}$ introduced 
in \cite{ohmi:1994},
\begin{equation}
\mbf{T}=\mbf{H}\mbf{V}\mbf{U}\mbf{V}^\dagger\mbf{H}^\dagger,
\label{ref:normal-def}
\end{equation}
where
\begin{align}
\mbf{U}&=\begin{pmatrix}
\mbf{U}_a & \mbf{0} & \mbf{0} \\
\mbf{0} & \mbf{U}_b & \mbf{0} \\
\mbf{0} & \mbf{0} & \mbf{U}_c
\end{pmatrix},\label{eqn:U-def}\\
\mathbf{V}&=\begin{pmatrix}
\mu\mathbf{I}_2 & \mathbf{V}_2 & 0\\
-\mathbf{V}_2^\dagger & \mu\mathbf{I}_2 & 0\\
0 & 0 & \mathbf{I}_2
\end{pmatrix}\label{eqn:V-def},\\
\mathbf{H}&=\begin{pmatrix}
\left(1-\frac{\left|\mathbf{H}_a\right|}{1+\rho}\right)\mathbf{I}_2 & 
-\frac{\mathbf{H}_a\mathbf{H}_b^\dagger}{1+\rho} & \mathbf{H}_a\\
-\frac{\mathbf{H}_b\mathbf{H}_a^\dagger}{1+\rho} & 
\left(1-\frac{\left|\mathbf{H}_b\right|}{1+\rho}\right)\mathbf{I}_2 &\mathbf{H}_b\\
-\mathbf{H}_a^\dagger & -\mathbf{H}_b^\dagger & \rho\mathbf{I}_2
\end{pmatrix}\label{eqn:H-def}.
\end{align}

The $2\times2$ submatrices are defined as,
\begin{equation}
\mathbf{V}_2=\begin{pmatrix}
v_{11} & v_{12}\\
v_{21} & v_{22}
\end{pmatrix},
\hspace{0.25cm}
\mathbf{H}_{a,b}=\begin{pmatrix}
\zeta_{a,b} & \eta_{a,b}\\
\zeta'_{a,b} & \eta'_{a,b}
\end{pmatrix},
\label{eqn:Hx-def}
\end{equation}
and $\mathbf{I}_2$ is the $2\times2$ identity matrix.  

$\mbf{U}$ describes the motion of the particles in normal-mode coordinates.
$\mbf{V}$ and $\mbf{H}$ describe how the normal-modes couple into
lab coordinates.

$\eta_{a,b}$ and $\eta'_{a,b}$ are referred to as normal-mode
dispersions and their derivatives.
$\zeta_{a,b}$ and $\zeta'_{a,b}$ are referred to as normal-mode
crabbing dispersions and their derivatives.
For the $a$-mode,
\begin{equation}
\mbf{U}_a=\begin{pmatrix}
\cos\theta_a+\alpha_a\sin\theta_a & \beta_a\sin\theta_a \\
-\gamma_a\sin\theta_a & \cos\theta_a-\alpha_a\sin\theta_a
\end{pmatrix},
\end{equation}
where $\theta_a$ is the phase advance per turn of the 
$a$-mode.
There are similar equations for $\mbf{U}_b$ and $\mbf{U}_c$.

For some $2n\times2n$ matrix $\mathbf{M}_{2n}$, 
the symplectic conjugate $\mathbf{M}_{2n}^\dagger$ is defined as,
\begin{equation}
\mathbf{M}^\dagger_{2n}=-\mathbf{S}_{2n}\mathbf{M}_{2n}^T\mathbf{S}_{2n},
\end{equation}
where $\mathbf{S}_{2n}$ is a $2n\times2n$ matrix whose $2\times2$ diagonal 
blocks are
\begin{equation}
\mathbf{S}_2=\begin{pmatrix}
0&1\\
-1&0
\end{pmatrix}.
\end{equation}
Note that if $\mathbf{M}_{2n}$ is symplectic, then 
$\mathbf{M}_{2n}^{-1}=\mathbf{M}_{2n}^\dagger$.
Superscript $^T$ denotes the matrix transpose.

\subsection{Computing the $6\times6$ parameterization}
In the general case, the normal-mode decomposition can be obtained via the 
eigen decomposition.  In \cite{wolski:2006}, the full turn matrix $\mbf{T}$ 
is decomposed into,
\begin{equation}
\mbf{T}=\mbf{A}\mbf{R}\mbf{A}^{-1},
\end{equation}
where $\mbf{A}$ and $\mbf{R}$ are real and symplectic.\footnote{$\mbf{A}$ here
is $\mbf{N}$ from Eq.~44 in Ref.~\cite{wolski:2006}.}
$\mbf{R}$ is block diagonal,
\begin{equation}
\mathbf{R}=\mathrm{diag}\left(\mathbf{R}_a,\mathbf{R}_b,\mathbf{R}_c\right),\\
\end{equation}
where
\begin{equation}
\mathbf{R}_{a,b,c}=\begin{pmatrix}
\cos\theta_{a,b,c} &\sin\theta_{a,b,c}\\
-\sin\theta_{a,b,c} &\cos\theta_{a,b,c}
\end{pmatrix}.
\end{equation}

$\mathbf{A}$ can be further decomposed to separate Twiss and coupling
information \cite{ohmi:1994},
\begin{equation}
\mathbf{A}=\mathbf{H}\mathbf{V}\mathbf{B}\mathbf{P}.
\label{eqn:NHVBP}
\end{equation}
$\mbf{H}$ and $\mbf{V}$ are the ``dispersion matrix'' and ``Teng matrix''
as defined above and $\mathbf{P}\mathbf{B}$ contains the Twiss information,
\begin{align}
\mbf{P}&=\mathrm{diag}
\left(\mathbf{P}_a,\mathbf{P}_b,\mathbf{P}_c\right)\label{eqn:P-def},\\
\mbf{P}_{a,b,c}&=\begin{pmatrix}
\cos\psi_{a,b,c} & \sin\psi_{a,b,c} \\
-\sin\psi_{a,b,c} & \cos\psi_{a,b,c}
\end{pmatrix},\\
\mbf{B}&=\mathrm{diag}
\left(\mathbf{B}_a,\mathbf{B}_b,\mathbf{B}_c\right)\label{eqn:B-def},\\
\mbf{B}_{a,b,c}&=\begin{pmatrix}
\sqrt{\beta_{a,b,c}} & 0\\
-\frac{\alpha_{a,b,c}}{\sqrt{\beta_{a,b,c}}} & \frac{1}{\sqrt{\beta_{a,b,c}}}
\end{pmatrix}.
\end{align}

$\mbf{P}$ is not an observable.
The three phases $\psi_a$, $\psi_b$, and $\psi_c$ are arbitrary and chosen
to give $\mbf{B}$ the desired form.

Equations~(\ref{eqn:V-def}), (\ref{eqn:H-def}), (\ref{eqn:P-def}), 
and (\ref{eqn:B-def}) are combined to write 
\begin{widetext}
\begin{equation}
\mbf{A}=\mbf{HVBP}=\begin{pmatrix}
\(\mu\(1-\frac{|\mbf{H}_a|}{1+\rho}\)\mbf{I}_2+
\frac{\mbf{H}_a\mbf{H}_b^\dagger\mbf{V}_2^\dagger}{1+\rho}\)\mbf{B}_a\mbf{P}_a & 
\(\(1-\frac{|\mbf{H}_a|}{1+\rho}\)\mbf{V}_2-
\mu\frac{\mbf{H}_a\mbf{H}_b^\dagger}{1+\rho}\)\mbf{B}_b\mbf{P}_b & 
\mbf{H}_a\mbf{B}_c\mbf{P}_c\\
-\(\(1-\frac{|\mbf{H}_b|}{1+\rho}\)\mbf{V}_2^\dagger+
\mu\frac{\mbf{H}_b\mbf{H}_a^\dagger}{1+\rho}\)\mbf{B}_a\mbf{P}_a & 
\(\mu\(1-\frac{|\mbf{H}_b|}{1+\rho}\)\mbf{I}_2-
\frac{\mbf{H}_b\mbf{H}_a^\dagger\mbf{V}_2}{1+\rho}\)\mbf{B}_b\mbf{P}_b & 
\mbf{H}_b\mbf{B}_c\mbf{P}_c\\
\(-\mu\mbf{H}_a^\dagger+\mbf{H}_b^\dagger\mbf{V}_2^\dagger\)\mbf{B}_a\mbf{B}_a &
\(-\mu\mbf{H}_b^\dagger-\mbf{H}_a^\dagger\mbf{V}_2\)\mbf{B}_b\mbf{B}_b &
\rho\mbf{B}_c\mbf{P}_c
\end{pmatrix}.
\label{eqn:HVBP}
\end{equation}
\end{widetext}
Because $|\mbf{B}_c\mbf{P}_c|=1$,
it is clear from Eq.~(\ref{eqn:HVBP}) that $\rho$ is the square root of 
the determinant of the lower right $2\times2$ block
of $\mbf{A}$.  It is then simple to obtain $\mbf{B}_c\mbf{P}_c$, as well 
as $\mbf{H}_a$ and $\mbf{H}_b$, which completely defines $\mbf{H}$.
This allows $\mathbf{V}\mathbf{B}\mathbf{P}$ to be obtained 
via $\mathbf{H}^\dagger\mathbf{A}$.  Similar steps then reveal $\mu$ 
and $\mathbf{V}_2$, defining $\mbf{V}$ and allowing $\mathbf{B}\mathbf{P}$ 
to be obtained.
$\psi_a$, $\psi_b$, and $\psi_c$ are selected to make the $\(1,2\)$, $\(3,4\)$,
and $\(5,6\)$
elements of $\mbf{V}^\dagger\mbf{H}^\dagger\mbf{A}\mbf{P}^\dagger$ zero.

Finally, the full turn matrix is written as,
\begin{equation}
\mbf{T}=\mbf{HVBPR}\mbf{P}^\dagger\mbf{B}^\dagger\mbf{V}^\dagger\mbf{H}^\dagger
\end{equation}
This eigen decomposition becomes a normal-mode
decomposition \cite{SaganRubin} by writing 
\begin{equation}
\mbf{U}=\mbf{BPR}\mbf{P}^\dagger\mbf{B}^\dagger,
\end{equation}
so that
\begin{equation}
\mbf{T}=\mbf{H}\mbf{V}\mbf{U}\mbf{V}^\dagger\mbf{H}^\dagger.
\end{equation}

\subsection{Projecting normal-mode coordinates into lab frame coordinates}
Lab frame coordinates $\mbf{x}=\(x,p_x,y,p_y,z,p_z\)$ and normal-mode 
coordinates $\mbf{a}=\(a,p_a,b,p_b,c,p_c\)$ are connected by
\begin{equation}
\mbf{x}=\mbf{HVa}.
\end{equation}
Writing out $\mbf{HV}$,
\begin{widetext}
\begin{equation}
\mbf{HV}=
\begin{pmatrix}
\mu\(1-\frac{|\mbf{H}_a|}{1+a}\)\mbf{I}_2+
\frac{\mbf{H}_a\mbf{H}_b^\dagger\mbf{V}_2^\dagger}{1+a} & 
\(1-\frac{|\mbf{H}_a|}{1+a}\)\mbf{V}_2-
\mu\frac{\mbf{H}_a\mbf{H}_b^\dagger}{1+a} & 
\mbf{H}_a\\
-\(1-\frac{|\mbf{H}_b|}{1+a}\)\mbf{V}_2^\dagger+
\mu\frac{\mbf{H}_b\mbf{H}_a^\dagger}{1+a} & 
\mu\(1-\frac{|\mbf{H}_b|}{1+a}\)\mbf{I}_2-
\frac{\mbf{H}_b\mbf{H}_a^\dagger\mbf{V}_2}{1+a} & 
\mbf{H}_b\\
-\mu\mbf{H}_a^\dagger+\mbf{H}_b^\dagger\mbf{V}_2^\dagger &
-\mu\mbf{H}_b^\dagger-\mbf{H}_a^\dagger\mbf{V}_2 &
a\mbf{I}_2
\end{pmatrix},
\end{equation}
\end{widetext}
we see that the $c$ normal mode is coupled into the
$x$ lab frame coordinate via $\mbf{H}_a$.

As shown in \cite{wolski:2006}, the eigen vectors of the full 
turn matrix $\mbf{M}$ are the same as the eigen vectors of
$\mbf{\Sigma S}$, where $\mbf{\Sigma}$ is the matrix of second
order moments of a Gaussian distribution matched to the machine lattice
functions.  
The $\mbf{\Sigma}$ matrix can be obtained from,
\begin{equation}
\mbf{\Sigma S}=\mbf{A}\mbf{D}\mbf{A}^{\dagger}
\label{eqn:sigma}
\end{equation}
where,
\begin{align}
\mbf{D}&=\begin{pmatrix}
0&\epsilon_a&0&0&0&0\\
-\epsilon_a&0&0&0&0&0\\
0&0&0&\epsilon_b&0&0\\
0&0&-\epsilon_b&0&0&0\\
0&0&0&0&0&\epsilon_c\\
0&0&0&0&-\epsilon_c&0
\end{pmatrix},
\end{align}
and $\epsilon_a$, $\epsilon_b$, and $\epsilon_c$ are the three
normal-mode emittances.  Note that normal-mode emittances
and eigen mode emittances are equivalent.  

Normalizing the $\mbf{H}$ and $\mbf{V}$
coupling matrices by the $\beta$-functions allows us to write
simplified expressions for the crabbing angles and 
beam sizes \cite{SaganRubin}.  We define,
\begin{align}
\overline{\mbf{H}}&=\mbf{B}^\dagger\mbf{H}\mbf{B}\label{eqn:Hbar}\\
\overline{\mbf{V}}&=\mbf{B}^\dagger\mbf{V}\mbf{B}\label{eqn:Vbar}.
\end{align}

The exact angle of the major axis of an ellipse in the $ij$ plane, where
$i$ and $j$ could be any of $x$, $p_x$, $y$, $p_y$, $z$, or $p_z$ is
\begin{equation}
\theta_{ij}=
\frac{1}{2}\textrm{arctan}\frac{2\sigma_{ij}}{\sigma_{ii}-\sigma_{jj}},
\label{eqn:angle}
\end{equation}
where $\sigma_{ij}$, $\sigma_{ii}$, and $\sigma_{jj}$ are elements of the 
beam $\mbf{\Sigma}$
matrix.

The tilt of the beam in some plane in terms of the $\mbf{HVBP}$ parameters 
can be found by combining 
Eqs.~(\ref{eqn:NHVBP}), (\ref{eqn:sigma}), (\ref{eqn:Hbar}), (\ref{eqn:Vbar}), 
and (\ref{eqn:angle}).

To first order in $\mbf{H}_x$, $\mbf{H}_y$, and $\mbf{V}_2$ the 
horizontal crabbing angle $\theta_{xz}$ and
vertical crabbing angle $\theta_{yz}$ are,
\begin{align}
\theta_{xz}&\approx\sqrt{\beta_a\beta_c}\frac{\epsilon_c\overline{\mbf{H}}_{15}-
\epsilon_a\overline{\mbf{H}}_{26}}{\epsilon_c\beta_c-\epsilon_a\beta_a},\\
\theta_{yz}&\approx\sqrt{\beta_b\beta_c}\frac{\epsilon_c\overline{\mbf{H}}_{35}-
\epsilon_b\overline{\mbf{H}}_{46}}{\epsilon_c\beta_c-\epsilon_b\beta_b}
\end{align}

In the limits $\epsilon_c\gg\epsilon_a$ and $\epsilon_c\gg\epsilon_b$,
\begin{align}
\theta_{xz}&\approx\sqrt{\frac{\beta_a}{\beta_c}}\overline{\mbf{H}}_{15}=
\zeta_a-\frac{\alpha_c}{\beta_c}\eta_a\label{eqn:thetaxz},\\
\theta_{yz}&\approx\sqrt{\frac{\beta_b}{\beta_c}}\overline{\mbf{H}}_{35}=
\zeta_b-\frac{\alpha_c}{\beta_c}\eta_b.
\end{align}
In CesrTA, $\epsilon_c\sim 10^{-6}$ m, $\epsilon_a\sim 10^{-9}$ m,
and $\epsilon_b\sim 10^{-12}$ m, so the approximation is valid.

The amplitude of $\alpha_c$ in a storage ring can be approximated 
by $\left|\tan\(\pi Q_s\)\right|$, where $Q_s$ is the 
synchrotron tune.  Because $Q_s$ is typically near zero,
$\beta_c$ is relatively constant and can be estimated by
$\beta_c\approx\frac{\alpha_p L}{\sin\nu_s}$.
Using values from Table~\ref{tab:cesrta}, 
the amplitude of $\alpha_c$ is $0.21$ and $\beta_c$ is $13.1$ m.

Equation (\ref{eqn:thetaxz}) shows that, to leading order, 
there are two contributions to the $xz$ tilt angle.  
It will be shown in Sec.~\ref{sec:simplified} that the first contribution 
can be generated by having non-zero dispersion in the rf cavities.  The other 
contribution is present in all storage rings and is proportional to the local 
dispersion and longitudinal Twiss $\alpha_c$.

The projection of the beam envelope into the lab frame is obtained
from the $\(1,1\)$, $\(3,3\)$, and $\(5,5\)$ elements of $\mbf{\Sigma}$.
These expressions are, in general, quite complicated.  To second
order in $\overline{\mbf{H}}_a$, $\overline{\mbf{H}}_b$, 
and $\overline{\mbf{V}}_2$ the beam sizes are,
\begin{widetext}
\begin{align}
\sigma_x^2&\approx\beta_a\(\(1-2\frac{|\overline{\mbf{H}}_a|}{1+\rho}\)\mu^2
\epsilon_a +\(1-2\frac{|\overline{\mbf{H}}_a|}{1+\rho}\)\(\overline{v}_{11}^2+
\overline{v}_{12}^2\)\epsilon_b+
\(\overline{\zeta}_a^2+\overline{\eta}_a^2\)\epsilon_c\),\\
\sigma_y^2&\approx\beta_b\(\(1-2\frac{|\overline{\mbf{H}}_b|}{1+\rho}\)
\mu^2\epsilon_b
+\(1-2\frac{|\overline{\mbf{H}}_b|}{1+\rho}\)\(\overline{v}_{12}^2+
\overline{v}_{22}^2\)\epsilon_a+\(\overline{\zeta}_b^2+
\overline{\eta}_b^2\)\epsilon_c\),\\
\sigma_z^2&\approx\beta_c\(a^2\epsilon_c+
\(\overline{\eta}_a^2+\overline{\eta}'_a{^2}\)\mu^2\epsilon_a+
\(\overline{\eta}_b^2+\overline{\eta}'_b{^2}\)\mu^2\epsilon_b\).
\end{align}
\end{widetext}

\begin{figure}
   \centering
   \includegraphics*[width=1.00\columnwidth]
                                 {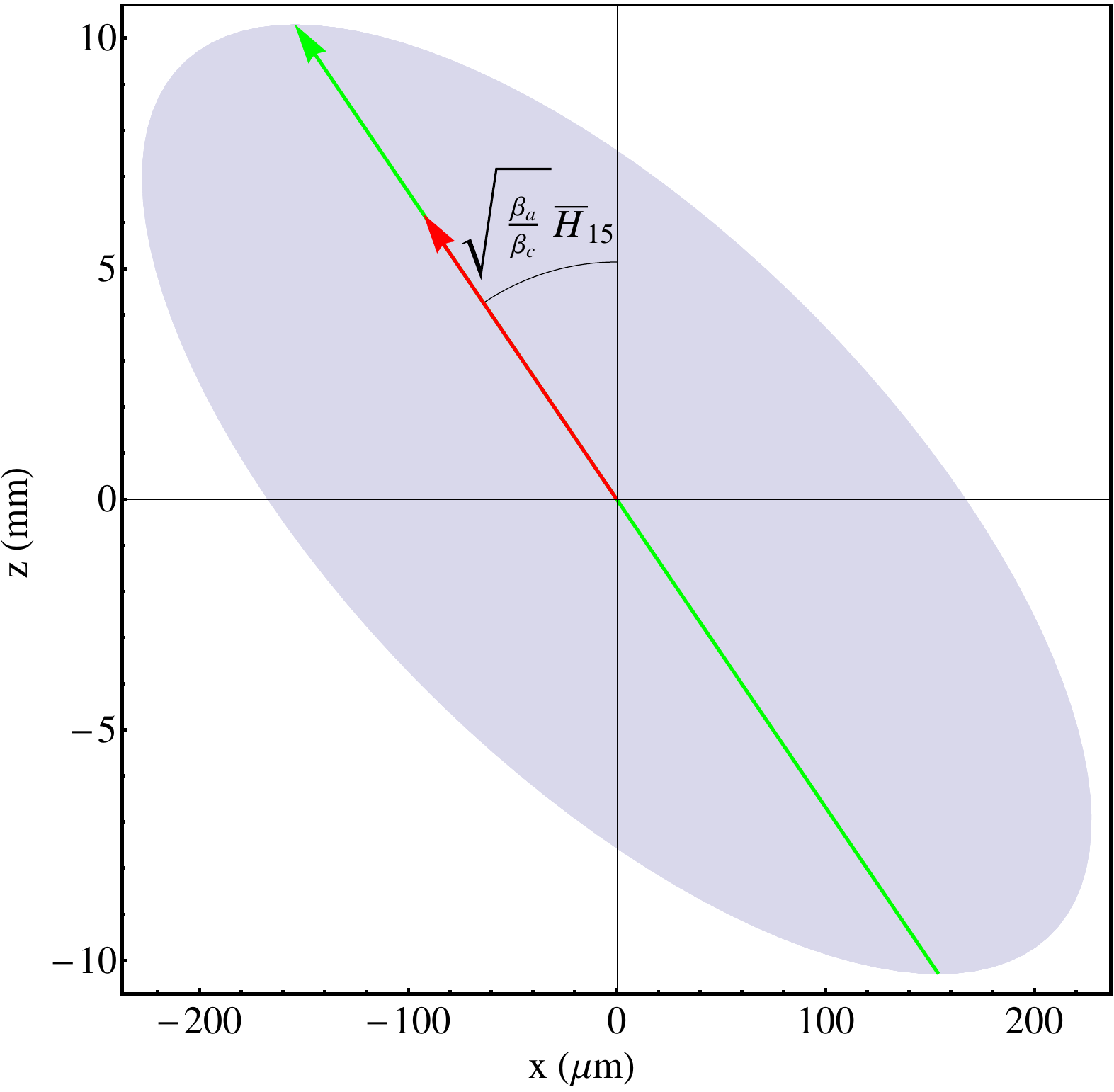}
   \caption{Projection of the beam envelop into the $xz$ plane with a crabbing
   angle of $-15$~mrad.  The green arrow is the major axis of the ellipse.
   The red arrow is crabbing angle obtained from the $\mbf{HVBP}$ 
   parameterization.  Note that the horizontal and vertical axes have very
   different scales in order to make the beam tilt discernible; a side effect
   is to make it appear as if the green arrow is not along the major axis.
   \label{fig:xz}}
\end{figure}

Figure~\ref{fig:xz} shows the $xz$ projection at the source point for the
horizontal beam size monitor.  The width of the $x$ projection of the beam
is $227$ $\mu$m.  Figure~\ref{fig:xz} also indicates the major axis of the
beam envelope and the tilt angle $\theta_{xz}$.

\begin{figure}
   \centering
   \includegraphics*[width=1.00\columnwidth]{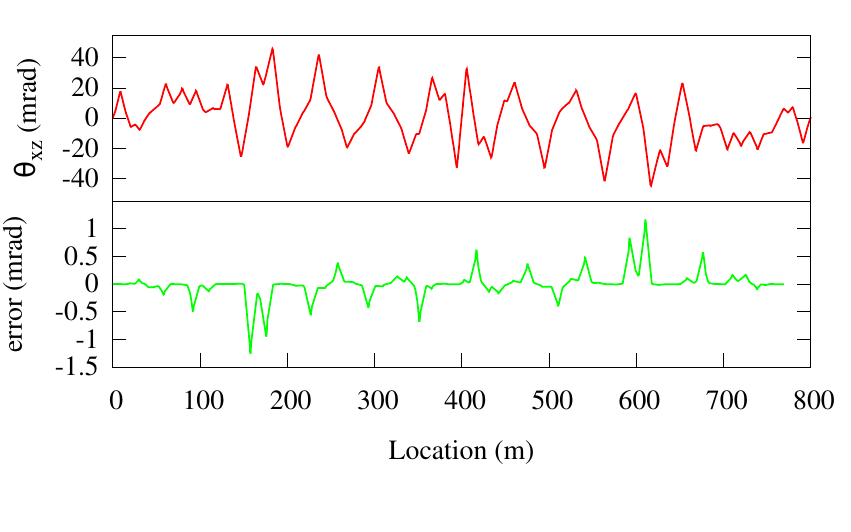} \vspace{-1.5cm}
   \caption{Top: $\theta_{xz}$ crabbing angle calculated 
   using Eq.~(\ref{eqn:thetaxz}).
   Bottom: difference between $\theta_{xz}$ obtained from 
   Eq.~(\ref{eqn:thetaxz}) and the actual 
   tilt of the beam envelope. \label{fig:xz-and-err}}
\end{figure}

Figure~\ref{fig:xz-and-err} shows $\theta_{xz}$ evaluated for a nominal 
CesrTA lattice using Eq.~(\ref{eqn:thetaxz}).

\subsection{$\mbf{H}_a$ as a function of cavity voltage in a simplified model}
\label{sec:simplified}
In the case where $xy$ and $yz$ coupling can be ignored, the normal-mode 
decomposition can be written in terms of the $2\times2$ blocks of the
$1$-turn matrix.
In this case, $\mbf{V}_2=\mbf{0}$, $\mbf{H}_b=\mbf{0}$, $\mu=1$.  
It is illuminating to consider this case because it demonstrates
how the presence of dispersion in the rf cavities generates beam tilt, and
it also suggests a method for mitigating the tilt in rings where there is 
tunable betatron phase advance between two or more sets of rf cavities.

The full turn matrix has the form
\begin{equation}
\mbf{T}=\begin{pmatrix}
\mbf{M} & \mbf{0} & \mbf{m} \\
\mbf{0} & \mbf{Y} & \mbf{0} \\
\mbf{n} & \mbf{0} & \mbf{N} \\
\end{pmatrix}\label{eqn:T-block},
\end{equation}
where $\mbf{M}$, $\mbf{m}$, $\mbf{Y}$, $\mbf{n}$, and $\mbf{N}$ are $2\times2$
matrices and $\mbf{0}$ is the $2\times2$ matrix of zeros.

The solution for $\mbf{H}$ is given by,
\begin{align}
a&=\sqrt{\frac{1}{2}+\frac{1}{2}\sqrt{\frac{\mathrm{Tr}\(\mbf{M}-\mbf{N}\)^2}
{\mathrm{Tr}\(\mbf{M}-\mbf{N}\)^2+4\left|\mbf{m}+\mbf{n}^\dagger\right|}}},
\label{eqn:a}\\
\mbf{H}_a&=\frac{\mbf{m}+\mbf{n}^\dagger}{a\sqrt{\mathrm{Tr}\(\mbf{M}-
\mbf{N}\)^2+4\left|\mbf{m}+\mbf{n}^\dagger\right|}}\label{eqn:Hx},
\end{align}
where $\mathrm{Tr}\(\cdot\)$ is the matrix trace.

The presence of horizontal dispersion in CesrTA's rf cavities 
introduces $xz$ and $p_xz$ coupling throughout the ring.  
This coupling causes $\zeta_a$ and $\zeta'_a$ to be non-zero.

To see how these coupling terms arise, consider a simplified model of a flat 
storage ring that is perfectly aligned
so there is no $xy$ nor $yz$ coupling.
The storage ring has an rf cavity located at Point $1$,
and we wish to determine $\mathbf{\zeta}_a$ at Point $0$.  The full turn matrix is
given by,
\begin{equation}
\mathbf{T}=\mathbf{T}_{10}\mathbf{T}_{RF}\mathbf{T}_{01},
\end{equation}
where $\mathbf{T}_{01}$ is the map from Point $0$ to Point $1$, 
and $\mathbf{T}_{RF}$ is the map for the rf cavity.  For simplicity, 
assume $\beta_x$ is uniform around the ring
and $\alpha_x=0$.  To obtain $\mbf{T}_{01}$,
populate Eq.~(\ref{eqn:T-block}) with 
\begin{equation}
\mathbf{M}_{01}=\begin{pmatrix}
\cos\Delta\phi_{01} & \beta_x\sin\Delta\phi_{01} \\
-\frac{1}{\beta_x}\sin\Delta\phi_{01} & 
\cos\Delta\phi_{01}
\end{pmatrix},
\end{equation}
where $\Delta\phi_{01}=\phi_1-\phi_0$ is the horizontal phase advance 
from $0$ to $1$, and
\begin{equation}
\mathbf{m}_{01}=\begin{pmatrix}
0 & \eta_{x1} \\
0 & \eta'_{x1}
\end{pmatrix}-\mathbf{M}_{01}\begin{pmatrix}
0 & \eta_{x0} \\
0 & \eta'_{x0}
\end{pmatrix},
\end{equation}
where $\eta_x$ and $\eta'_x$ are the ordinary dispersion and its derivative,  i.e.
the dispersion in the limit of zero rf voltage.
Also define,
\begin{equation}
\mathbf{N}_{01}=\begin{pmatrix}
1 & L_{01}\alpha_{01}^p \\
0 & 1
\end{pmatrix},
\end{equation}
where $L_{01}$ and $\alpha_{01}^p$ are the fraction of the total circumference 
from $0$ to $1$ and effective momentum compaction
between $0$ and $1$, respectively.
From the symplecticity of $\mathbf{M}$ we have,
\begin{equation}
\mathbf{n}_{01}^T=\mathbf{M}_{01}^T
\mathbf{S}_2\mathbf{m}_{01}\mathbf{N}_{01}^{-1}\mathbf{S}_2.
\end{equation}
Because the ring is flat and ideal, $\mbf{Y}_{01}=\mbf{I}_2$.

The transfer matrix $\mbf{T}_{\mathrm{rf}}$ for the rf cavity is given by
\begin{align}
\mbf{M}_{\mathrm{rf}}&=\mbf{Y}_{\mathrm{rf}}=\mbf{I}_2\\
\mbf{m}_{\mathrm{rf}}&=\mbf{n}_{\mathrm{rf}}=\mbf{0}\\
\mbf{N}_{\mathrm{rf}}&=\begin{pmatrix}
1 & 0\\
\tilde{V} & 1
\end{pmatrix}
\end{align}
where $\tilde{V}$ was defined in Eq.~(\ref{eqn:Vtilde}).

The synchrotron tune as a function of $\tilde{V}$ is,
\begin{equation}
Q_s=\frac{1}{2\pi}\sqrt{\frac{-L\eta_\gamma\tilde{V}}{4\pi^2\beta_r^2}\cos\Psi_0},
\end{equation}
where $\eta_\gamma$ is the slip factor.

Evaluating Eqs.~(\ref{eqn:a}) and (\ref{eqn:Hx}) for 
$\mbf{T}_{10}\mbf{T}_{RF}\mbf{T}_{01}$ 
yields,
\begin{widetext}
\begin{align}
\zeta_{a0} =&\frac{\tilde{V}}{a\chi}\Big(\eta_{x1}\(\cos\Delta\phi_{01}-
\cos\(\nu_x-\Delta\phi_{01}\)\)
-\beta_x\eta'_{x1}\(\sin\Delta\phi_{01}+\sin\(\nu_x-\Delta\phi_{01}\)\)\Big)\label{eqn:zeta-a0}\\
\eta_{a0} =&\frac{2}{a\chi}\(1-\cos\nu_x\)\eta_{x0}-\frac{\tilde{V}}{a\chi}\Big(
\(L_{01}\alpha^p_{01}+L_{10}\alpha^p_{10}\)\eta_{x0}+
\(L_{10}\alpha^p_{10}\cos\Delta\phi_{01}+
L_{01}\alpha^p_{01}\cos\Delta\phi_{10}\)\eta_{x1}-\nonumber\\
&\(L_{10}\alpha^p_{10}\sin\Delta\phi_{01}-
L_{01}\alpha^p_{01}\sin\Delta\phi_{10}\)\beta_x\eta'_{x1}
\Big)\label{eqn:eta-a0}
\end{align}
\end{widetext}
where $\chi=\sqrt{\textrm{Tr}\left(\mathbf{M}-\mathbf{N}\right)^2+ 
4\left|\mathbf{m}+\mathbf{n}^\dagger\right|}$.

We find that as anticipated $\zeta_{a0}$ is proportional to the dispersion
in the cavity and the accelerating voltage.

In the absence of $xz$ coupling, 
$\left|\mbf{m}+\mbf{n}^\dagger\right|=0$.  
In that case, $\chi=2\cos\nu_x-2\cos\nu_s$ and $a=1$.
As rf voltage goes to zero, which implies a 
small $\nu_s$, the normal-mode dispersion $\eta_{a0}$ becomes the 
ordinary dispersion $\eta_{x0}$.

\section{LATTICE DESIGN}
\label{sec:lats}
In a storage ring with multiple sets of rf cavities, $\zeta_{a0}$ 
and $\eta_{a0}$ at the instrumentation source point are given by the sum of 
the contributions from each cavity.  

For a given pair of cavities, assume that the coupling is small
such that $\left|\mbf{m}+\mbf{n}^\dagger\right|$ is small.  Then
$\chi_1\approx\chi_2$, where subscripts $1$ and $2$ denote the
separate cavities.  Further assuming that $\eta_1=\eta_2$
and $\eta'_1=\eta'_2$, we find that 
\begin{equation}
\sum\zeta_{a0}\propto\cos{\frac{\Delta\phi_{12}}{2}},
\end{equation}
where $\Delta\phi_{12}$ is the betatron phase advance between
the cavities.  A phase advance of $\(2n+1\)\frac{\pi}{2}$
causes the $\zeta_0$ generated in one cavity to cancel out
the $\zeta_0$ generated in the other cavity.  

In the case of a ring with only one rf cavity, or if the
$\(2n+1\)\frac{\pi}{2}$ condition cannot be met, $\zeta_a$ can
be minimized at the instrumentation source point by adjusting the 
betatron phase advance such that the tilt passes through a zero
at the instrumentation source point.

In CESR, there are two pairs of rf 
cavities, separated by about $1.5$ betatron wavelengths.  Because
$\eta_1,\eta'_1$, and $\chi_1$ are only approximately equal to
$\eta_2,\eta'_2$, and $\chi_2$, $\zeta_a$ at the observation point
is minimized using a optimizer that varies quadrupole strengths.
The minimization procedure results in an approximately closed $xz$ 
coupling bump through the south region of the accelerator.

\begin{figure}
   \centering
   \includegraphics*[width=1.00\columnwidth]{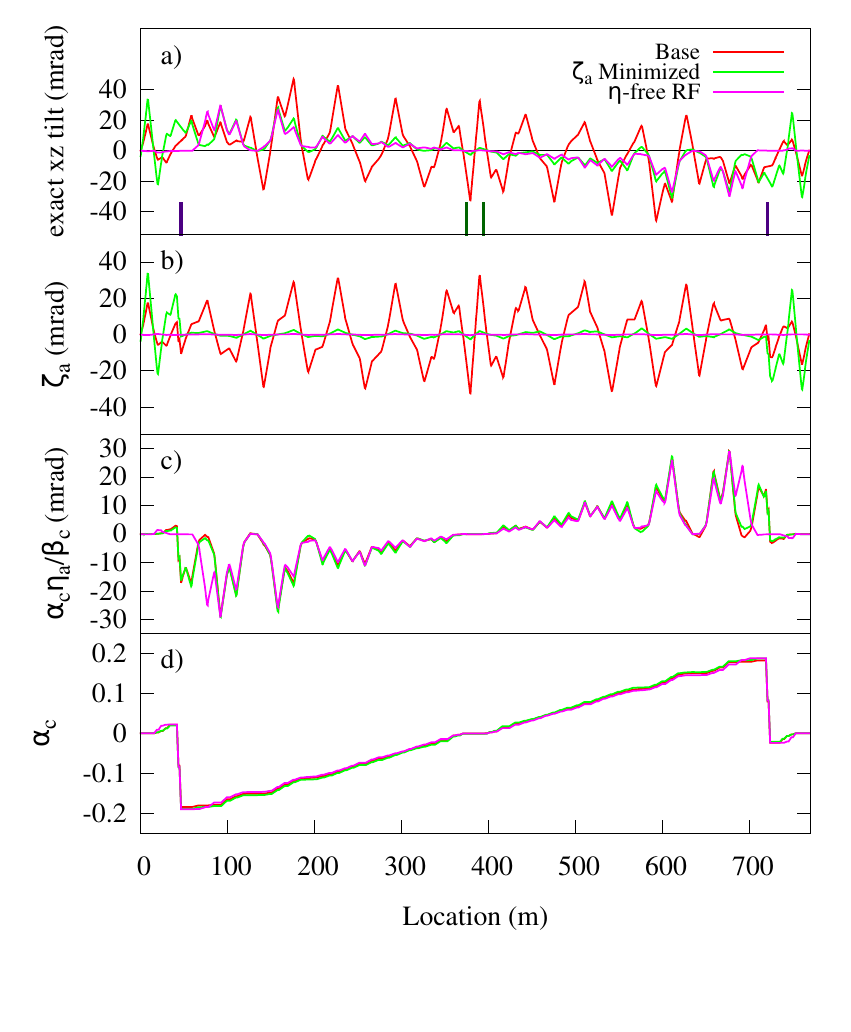} \vspace{-1.5cm}
   \caption{For lattices with $6.3$ MV total rf voltage: 
   (a) Tilt of beam in $xz$ plane calculated from major axis
   of projected beam envelope.  Purple bars indicate the location of the two
   pairs of RF cavities.  Green bars indicate the location of horizontal beam
   size monitors.  (b) Crabbing contribution from 1st term 
   in Eq.~(\ref{eqn:thetaxz}).  (c) Crabbing contribution from 
   2nd term in Eq.~(\ref{eqn:thetaxz}).
   (d) Longitudinal Twiss parameter $\alpha_c$.\label{fig:zeta}}
\end{figure}

As shown in Eq.~(\ref{eqn:thetaxz}), the beam tilt has a contribution
from $\alpha_c$, in addition to $\xi_a$.  As shown in Fig.~\ref{fig:zeta},
$\alpha_c$ is naturally zero at the horizontal beam size measurement
source point for CesrTA.

A second method to eliminate $\zeta_a$ is to zero the horizontal dispersion 
in the rf cavities.  In CESR this forces non-zero dispersion in the nearby 
damping wiggler straight, and therefore requires that $6$ of the $12$ 
wigglers be powered off.  In Fig.~\ref{fig:cesrta}, these are the South end
wiggler triplets at six o'clock.

Figure~\ref{fig:zeta} shows the model $\zeta_a$ values for the base lattice,
the lattice with $\zeta_a$ minimized, and a lattice with zero dispersion in 
the rf cavities.  ``Base'' refers to the base CesrTA lattice.  This lattice 
has $\sim1$ m horizontal dispersion in each pair of rf cavities.
``$\zeta_a$ minimized'' refers to the lattice optimized to minimize the effect
of crabbing at the instrumentation source points.
``$\eta$ free'' refers to the lattice with zero dispersion in the rf cavities.  

The zero current horizontal and vertical emittances of the ``$\eta$ free'' 
lattice are larger because $6$ of the $12$ damping wigglers are powered off.  
The damping time of the ``Base'', ``$\zeta_a$ minimized'', 
and ``$\eta$ free'' lattices are $56.6$ ms, $56.6$ ms, and $99.8$ ms, 
respectively.

\section{EXPERIMENT}\label{sec:experiment}
Measurements are taken with each of the three lattices at $2.1$ GeV using a 
single bunch of positrons.  The experiment is conducted by setting the rf 
voltage, then taking several bunch size measurements at both $0.5$ 
and $1.0$ mA.  Horizontal and vertical beam size and bunch length are 
recorded from $6.3$ MV down to $1.0$ MV in roughly $1$ MV increments.
The total rf voltage is split roughly equally among the four rf cavities.

Vertical beam size is measured by imaging x rays from a hard bend onto
a vertical diode detector array \cite{rider:2012,xbsm:handbook}.  
Horizontal beam size is measured using an interferometer
which images synchrotron radiation from a soft bend \cite{suntao:2012}.  
Bunch length measurements are made
with a streak camera using synchrotron radiation from the same bend 
\cite{holtzapple:2000}.

The simulation includes intrabeam scattering (IBS) calculated
using the Kubo-Oide formalism \cite{kubooide:2001}.  IBS occurs when 
collisions among the particles that compose a beam transfer momentum 
between the particles such that the emittance of the beam is changed. 
The implementation of this formalism at CesrTA is discussed in 
\cite{prst_ibs}.

\begin{figure}
   \centering
   \subfloat{
   \includegraphics*[width=0.93\columnwidth]{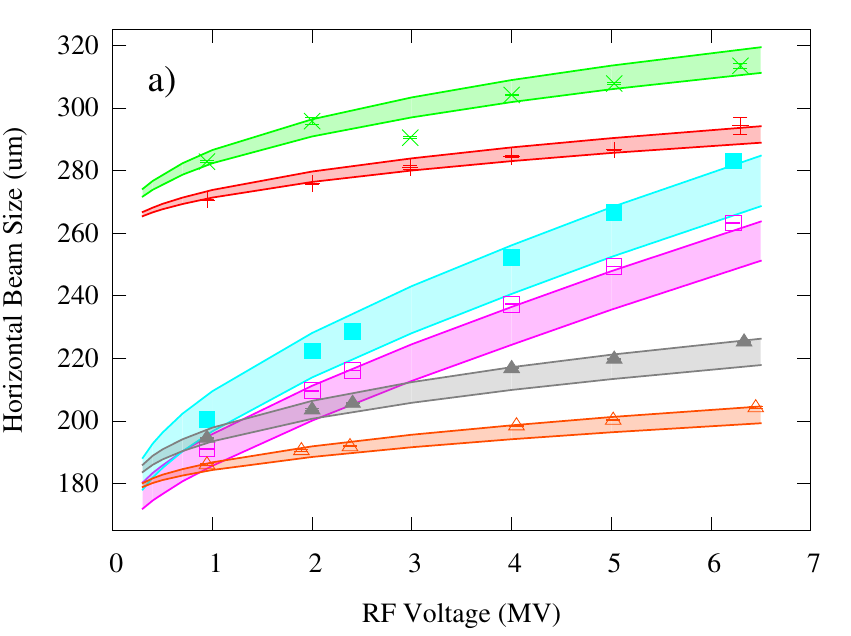}
   \label{fig:v15-h}}\\ \vspace{-0.4cm}
   \subfloat{
   \includegraphics*[width=0.93\columnwidth]{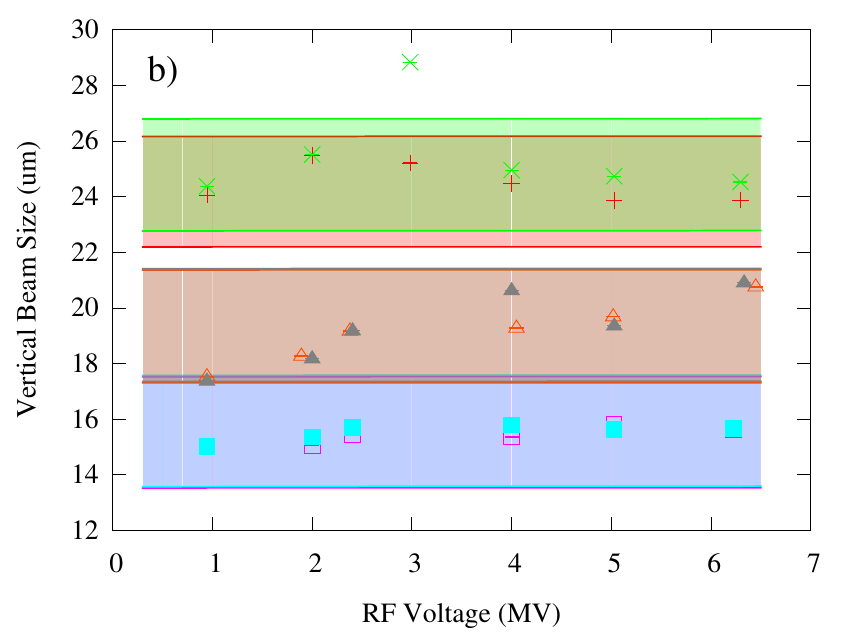}
   \label{fig:v15-v}}\\ \vspace{-0.4cm}
   \subfloat{
   \includegraphics*[width=0.93\columnwidth]{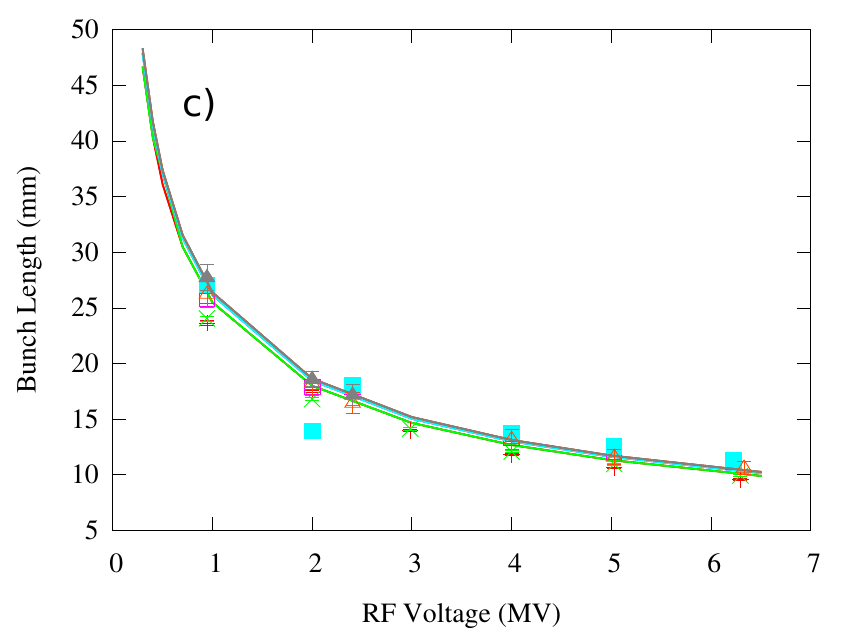}
   \label{fig:v15-l}}\\ \vspace{-0.4cm}
   \subfloat{
   \includegraphics*[width=0.50\columnwidth]{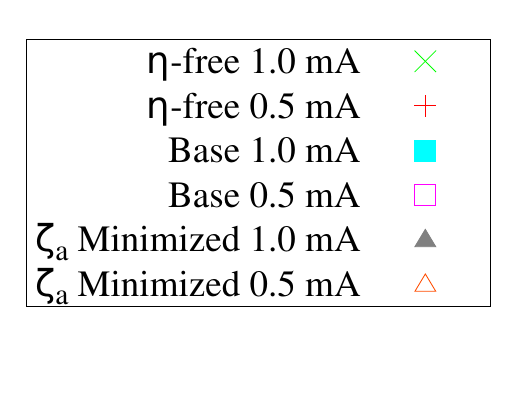}
   \label{fig:v15-key}} \vspace{-0.6cm}
   \caption{\protect\subref{fig:v15-h} Horizontal, \protect\subref{fig:v15-v} 
   vertical, and \protect\subref{fig:v15-l} bunch length measurements and 
   simulation results.  Points are data and curves are simulation results.  
   The data points are the average over several measurements and the error 
   bars are the statistical uncertainty. In CesrTA, 
   $1.0$ mA $= 1.6\times10^{10}$ particles/bunch.\label{fig:v15}}
\end{figure}

Figure~\ref{fig:v15} shows the measured horizontal, vertical, and
longitudinal bunch size, along with simulation results.  
Beam sizes are obtained from the simulation by
projecting the beam envelope into lab frame coordinates
using Eq.~(\ref{eqn:sigma}).

The simulation takes as parameters the zero current $a$-mode
and $b$-mode emittances.  These emittances are obtained by fitting the data.

The vertical beam size measurement is subject to a $\pm2$ $\mu$m systematic
uncertainty.  This is accounted for by running the simulation twice,
adjusting the $b$-mode emittance parameter to span a $4$ $\mu$m range
of vertical beam size.  This parameter range is the source of the colored
bands in the plots.

The horizontal beam size in the ``Base'' lattice has an additional source
of systematic uncertainty due to the beam tilt.
The dependence of the projected horizontal beam size $\sigma_x$ 
on the $xz$ tilt angle $\theta_{xz}$ is
\begin{equation}
\frac{d\sigma_{x}}{d\theta_{xz}}=\frac{\theta_{xz}\sigma_z^2}{\sigma_x},
\end{equation}
where $\sigma_z$ is the bunch length.
At the horizontal beam size monitor source point, 
\begin{align}
\frac{d\sigma_{x}}{d\theta_{xz}}&=\phantom{-}6.6 \frac{\mu \mathrm{m}}{\mathrm{mrad}},\\
\frac{d\theta_{xz}}{ds}&=-1.4 \frac{\mathrm{mrad}}{\mathrm{m}}.
\end{align}
The uncertainty in the location of the source point, given by
the depth of field, is $0.35$ m \cite{suntao:2012}.
This contributes an additional $\pm3.2$ $\mu$m systematic uncertainty
to the ``Base'' lattice simulation results.

For the ``$\zeta_a$ Minimized'' lattice, the same zero current
$a$-mode emittance and $b$-mode emittance is used for all data points.
This is true also for the ``$\eta$-free'' lattice.

The presence of $\zeta_a$ in a storage ring affects the zero current
$a$-mode emittance.  The dependence of the zero current emittance 
on rf voltage is estimated using PTC \cite{forest}.  This dependence is 
significant only for the ``Base'' lattice.

Table \ref{tab:emits} shows the emittances used to generate the 
simulation results.

\begin{table}[tb]
\centering
\caption{Simulation parameters: zero current emittances.  The presence
of $\zeta_a$ in the ``Base'' lattice creates an rf voltage dependent 
zero current $a$-mode emittance.\label{tab:emits}}
\begin{tabular*}{\columnwidth}{@{\extracolsep{\fill}}lccc}
\hline
\hline
Lattice & $V_{\mathrm{rf}}$  & $\epsilon_a$   & $\epsilon_b$ \\
\hline
$\eta$-free & all          & $5.2$ nm & $12.3$ pm - $17.9$ pm\\
$\zeta_a$ Minimized & all  & $3.3$ nm & $7.3$ pm - $11.1$ pm\\
\multirow{7}{*}{Base}  & $<1.0$ MV & $2.99$ nm  & 
                                  \multirow{7}{*}{$4.58$ pm - $7.69$ pm}\\
                       & $1.0$ MV  & $3.00$ nm &\\
                       & $2.0$ MV  & $3.02$ nm &\\
                       & $3.0$ MV  & $3.05$ nm &\\
                       & $4.0$ MV  & $3.08$ nm &\\
                       & $5.0$ MV  & $3.13$ nm &\\
                       & $6.5$ MV  & $3.21$ nm &\\
\hline
\hline
\end{tabular*}
\end{table}

The differences in the measurement results between $0.5$ and $1.0$ mA are due 
to IBS.  The effect is most noticeable in the horizontal (Fig.~\ref{fig:v15-h})
due to the large amount of horizontal dispersion throughout the ring.  The rms 
horizontal dispersion is $1$ m.

The jump in vertical beam size at $\sim 3$ MV for the ``$\eta$ free'' lattice 
at $1.0$ mA (Fig.~\ref{fig:v15-v}) was due to crossing a synchrobetatron 
resonance.

The model lattices used for the simulation
are ideal, with no vertical dispersion or transverse coupling.
The result is that the simulation predicts negligible IBS blow up in the 
vertical dimension.  The insensitivity of the measured vertical beam size to 
changes in current and bunch length for the ``Base'' and ``$\eta$ free'' 
lattices suggest that transverse coupling and vertical dispersion are 
well-corrected.  The vertical dispersion is measured to be less than $15$ mm. 
The coupling is measured using an extended Edwards-Teng formalism 
to be $\overline{V}_{12}<0.003$.

An upward trend is suggested in the vertical beam size for 
the ``$\zeta_a$ minimized'' lattice (Fig.~\ref{fig:v15-v}).  It is unlikely 
that this trend is due to IBS, as the beam size is the same for the $0.5$ 
and $1.0$ mA data points.  Such a trend could arise from an optics error 
introducing $yz$ coupling.

\section{CONCLUSION}\label{sec:conclusion}
The decomposition of the $1$-turn matrix into the coupling 
matrices $\mbf{H}$ and $\mbf{V}$ and Twiss matrix $\mbf{B}$ yields
useful information about the coupling properties of a ring.  We showed
(Eq.~\ref{eqn:thetaxz}) that the beam has an $xz$ tilt given by the
crabbing dispersion $\zeta_a$ and a term that is proportional to the local 
horizontal dispersion.  We showed how the presence of dispersion
in the rf cavities can generate crabbing dispersion throughout the ring.

The beam size measurements versus rf voltage
from the ``$\eta$ free'' and ``$\zeta_a$ minimized'' lattices agree well 
with simulation.  The dependence of the measured horizontal beam size on
rf voltage is due entirely to IBS effects.  The residual tilt
after correction is negligible.  
Evidently, our methods for mitigating $xz$ tilt are effective.

For the ``Base'' lattice, which has significant $\zeta_a$ throughout, 
$xz$ tilt makes a nonnegligible contribution to projected horizontal
beam size.  Agreement between the ``Base'' lattice beam size measurements 
and simulation result is reasonable and suggests that the projected size of 
coupled beams can be reliably computed using Eq.~(\ref{eqn:sigma}).

As described in Sec.~\ref{sec:intro}, crabbing affects beam size 
measurements and has potential applications in mitigating the 
effects of angle crossing in a collider, and in generating 
subpicosecond pulses in a light source.
In Sec.~\ref{sec:toy} and Sec.~\ref{sec:theory} we explained through 
two different methods,
how dispersion in ordinary rf cavities generates crabbing and how
the tilt angle of the beam can be calculated.
In Sec.~\ref{sec:lats} we presented methods for controlling, or mitigating,
the amount of tilt.  This was followed by Sec.~\ref{sec:experiment}, 
where we supported our theory with experiment.

\begin{acknowledgments}
The experiments reported here
would not have been possible without the diligent support of the
CESR Operations Group.

This research was supported by
NSF and DOE contracts PHY-0734867, PHY-1002467,
PHYS-1068662, DE-FC02-08ER41538, and DE-SC0006505.
\end{acknowledgments}

\bibliography{prst_v15}

\end{document}